\documentclass[11pt,a4paper]{article}
\usepackage{graphicx}
\usepackage{jcappub}
\usepackage{subfigure}

\def\apj{ApJ}%
\def\apjl{ApJ}%
\def\apjs{ApJS}%
\def\apss{Ap\&SS}%
\def\aap{A\&A}%
\def\aapr{A\&A~Rev.}%
\def\jcap{J. Cosmology Astropart. Phys.}%
\def\mnras{MNRAS}%
\def\na{New A}%
\def\nat{Nature}%
\title{First axion bounds from
a pulsating helium-rich white dwarf star}

\author[a,b]{T. Battich,}
\author[a,b]{A. H. C\'orsico,}
\author[a,b]{L. G. Althaus,}
\author[b,c]{M. M. Miller Bertolami,}

\affiliation[a]{Grupo de Evoluci\'on Estelar y Pulsaciones. 
                Facultad de Ciencias Astron\'omicas y Geof\'{\i}sicas,
                Universidad  Nacional de La Plata,
                Paseo del  Bosque s/n,
               (1900) La Plata,
                Argentina}
\affiliation[b]{Instituto de Astrof\'{\i}sica La Plata,
                CONICET-UNLP,
                Argentina}
\affiliation[c]{Max-Planck-Institut f\"ur Astrophysik,
                Karl-Schwarzschild-Str. 1, 8574, Garching
                Germany}

\emailAdd{tbattich@fcaglp.unlp.edu.ar}

\abstract{The Peccei-Quinn mechanism proposed to solve the CP problem of Quantum Chromodynamics has as consequence the existence of axions, hypothetical weakly interacting particles whose mass is constrained to be on the sub-eV range. If these particles exist and interact with electrons, they would be emitted from the dense interior of white dwarfs, becoming an important energy sink for the star. Due to their well known physics, white dwarfs are good laboratories to study the properties of fundamental particles such as the axions.
We study the general effect of axion emission on the evolution of helium-rich white dwarfs and on their pulsational properties. 
To this aim, we calculate evolutionary helium-rich white dwarf models with axion emission, and asses the pulsational properties of this models. Our results indicate that the rates of change of pulsation periods are significantly affected by the existence of axions. 
We are able for the first time to independently constrain the mass of the axion from the study of pulsating helium-rich white dwarfs. To do this, we use an estimation of the rate of change of period of the pulsating white dwarf PG 1351+489 corresponding to the dominant pulsation period. 
From an asteroseismological model of PG 1351+489 we obtain $g_{ae}<3.3\times10^{-13}$ for the axion-electron coupling constant, or $m_a\cos^2{\beta}\lesssim$ 11.5 meV for the axion mass. This constraint is relaxed to $g_{ae}<5.5\times10^{-13}$ ($m_a\cos^2{\beta}\lesssim$ 19.5 meV), when no detailed asteroseismological model is adopted for the comparison with observations.}

\notoc

  
\begin{document}

\maketitle  


\section{Introduction}
\label{introduction}
One of the main unsolved problems of the standard model of particle physics is the so-called CP-problem of Quantum Chromodynamics. The Standard Model does not provide a reason for the absence of CP-symmetry violation in strong interactions. In 1977, \cite{peccei} proposed a mechanism to solve this problem. One of the consequences of this mechanism would be the existence of a new particle, the so-called {\it axion} \cite{weinberg}. The proposal of \cite{peccei} in its original form was ruled out by the experiments (e.g. \cite{1987NuPhB.279..401B} and references there in). 
Currently, there are being discussed two types of axion models that are called {\it invisible axion models}, such as the KSVZ model \cite{kim:1979,shifman:1980} and the DFSZ model \cite{dine:1981,Zhitnitsky:1980tq}. In the KVSZ model axions couple to hadrons and photons, while in the DFSZ model these particles also couple to charged leptons like electrons.  
In this work we are interested in DFSZ axions, i.e. those that interact with electrons.

The coupling strength of DFSZ-type axions to electrons is defined by an adimensional coupling constant $g_{ae}$, whose expression is:
\begin{equation}
  g_{ae}=2.8\times 10^{-14} \, \frac{m_{a}\cos^2{\beta}}{1 \mathrm{meV}}
\end{equation}
 where $m_{a}$ is the axion mass in units of meV, and $\cos^2{\beta}$ is a free, model dependent parameter. Axions have been proposed 
 as candidates for dark matter \cite{2007JPhA...40.6607R}. The contribution of these particles to dark matter is dependent on their mass (or coupling constant). However, the models do not provide information about the value of the axion mass, and therefore it has to be inferred from observations. 
In particular, stars can be employed to place constraints on the mass of the axion \cite{raffelt:book:1996,2007JPhA...40.6607R}. 
 DFSZ-type axions would be emitted from the dense core of white dwarfs (WDs), which are among the most studied classes of stars. These stars represent the final stage of the evolution of low and intermediate mass stars (see \cite{review} and references therein). 
Our knowledge of the physical processes governing the evolution of WDs relies on solid grounds, since the basic principle in their evolution is a well understood and simple cooling process. This fact makes WDs useful laboratories to provide constraints on  
$g_{ae}$ \cite{1986PhLB..166..402R}. Due to their large mean free path, if axions of DFSZ type exist, would be emitted and would scape almost freely from the WD interiors. Therefore, the emission of axions would increase the cooling rate of a WD. For increasing values of axion mass, and hence of the coupling constant $g_{ae}$, the axion emission becomes higher, thus implying an important energy sink and larger cooling rates for the WD. 

As WD stars evolve, they pass through stages where they are pulsationally unstable. Some of the observed WDs are pulsating stars, that exhibit luminosity variations with periods between 100-1500$\,$s due to non-radial $g$ (gravity) modes, which are a type of modes whose main restoring force is gravity. 
The pulsation periods of variable WDs change secularly due to the adjustments in the mechanical and thermal structure throughout their evolution. 
The rate at which periods vary depends on two independent processes acting together \cite{winget}: 
the cooling of the region where the oscillations are relevant, and the gravitational contraction of the outer layers of the star. The first process has the consequence of increasing the periods, while the second one has the opposite effect, shifting the periods to shorter values. This last process is less important than the former in the case of DAV and DBV WDs (pulsating WDs with hydrogen-rich and helium-rich surface compositions, respectively), where the degeneracy degree of the plasma prevents the star to have significant changes on its radius. Then, it is expected that DAV and DBV WDs have positive values of the rate of change of their periods ($\dot \Pi$). 
If the rate of cooling of the star increases, the lengthening of the periods became greater. 
Therefore, the existence of axions would have the effect of increase $\dot \Pi$, the magnitude of this increase being larger for a grater axion coupling constant. In this sense, measurements of $\dot \Pi$ in pulsating WDs can be used to put bounds to the axion-electron coupling constant. 
The possibility of using the rate of change of period of a pulsating WD to constrain the value of the axion mass was first proposed by \cite{isern}. 
They obtained a value of $m_a \cos^2\beta\lesssim 8.7$ meV ($g_{ae}\lesssim 2.4 \times 10^{-13}$) using a measurement of $\dot \Pi$
corresponding to the mode of period 215 s of the DAV G117-B15A. 
\cite{ale0} derived a bound to the axion mass of $m_a \cos^2 \beta\lesssim 4.4$ meV ($g_{ae}\lesssim 1.2 \times 10^{-13}$) using, for the first time, an asteroseismological model for G117-B15A and a new measurement of the $\dot \Pi$ of the same period of pulsation. Later, \cite{bischoff:kim:2008} obtained an upper limit of 13.5 meV ($g_{ae}\lesssim 3.8 \times 10^{-13}$) or 26.5 meV ($g_{ae}\lesssim 7.4 \times 10^{-13}$) depending on the thickness of the hydrogen (H) envelope of the asteroseismological model. The measured value of $\dot \Pi$ corresponding to the mode with period of 215 s of G117-B15A has varied with time, approaching in the last measurements to a value of $\dot \Pi \sim 4 \times 10^{-15}$ s/s \cite{2010A&A...512A..86I}. Using the latest measure of $\dot \Pi$ and a new detailed asteroseismological model developed by \cite{2012MNRAS.420.1462R}, \cite{ale} derived a new value for the mass of the axion. 
These authors report that the difference between the observed and theoretical rates of change of period is consistent with the emission of axions with $m_a \cos^2 \beta = 17.4^{+2.3}_{-2.7}$ ($g_{ae}\sim 4.9^{+0.6}_{-0.8} \times 10^{-13}$). \cite{ale2} obtained a similar result using a measurement of the rate of change of period of another DAV star, R548, corresponding to the pulsation mode with a period of 213 s. In their analysis they obtained a value of $m_a \cos^2 \beta = 17.1^{+4.3}_{-5.8}$ ($g_{ae}\sim 4.8^{+1.2}_{-1.6} \times 10^{-13}$).

Besides the measurement of the secular period drift of pulsating WDs, WD cooling times can also be inferred from the luminosity function of white dwarfs (WDLF). 
 Comparing the WDLFs constructed from theoretical models with observed WDLFs, it is possible to constrain the value of the axion mass. This technique was first used by \cite{isern:2008,isern:2009}, who employed the Galactic Disk WDLF. These authors found  
that values greater than 10 meV for the axion mass should be discarded. \cite{miller:2014} revisited the effects of axion emission in the Galactic WDLF. 
carried out in that work disfavour axion masses in the range $m_a \cos^2 \beta \gtrsim 10$ meV ($g_{ae} \gtrsim 2.8 \times 10^{-13}$) while lower axion masses cannot be discarded.   
\cite{2015ApJ...809..141H} derived a limit to the axion mass from the hot WDLF of the globular cluster 47 Tucanae. These authors claim a 95\% confidence limit (CL) of $g_{ae} < 8.4 \times 10^{-14}$, which implies a tight constraint on the axion mass of $m_a < 3$ meV. 
Another bound to the axion mass can be obtained from the ignition of helium in low-mass stars in the red giant branch. In particular, \cite{raffelt:1995} obtained an upper bound of $ m_{a} \cos^2\beta\lesssim 9$ meV ($g_{ae} \lesssim 2.5 \times 10^{-13}$). More recently, \cite{2013PhRvL.111w1301V} obtained a constraint at  68\% CL of  $m_a \cos^2\beta < 9.3$ meV ($g_{ae} < 2.6 \times 10^{-13}$). Finally, the search for solar axions from the Korea Invisible Mass Search experiment at the Yangyang Underground Laboratory gives as a result, for DFSZ axions, a limit at 90\% CL of $m_a \cos^2\beta < 496$ meV ($g_{ae} < 1.39 \times 10^{-11}$) \cite{Yoon:2016ogs}.

The constraints mentioned above do not agree (at 68\% CL) with the last values of the axion mass derived from the asteroseismological studies of the DAV G117-B15A \cite{ale} and R548 \cite{ale2}. It is worth mentioning that G117-B15A and R548 have similar structural and pulsational features. Both stars have comparable effective temperatures and surface gravities, and also the pulsation modes for which it has been possible to measure a value of $\dot \Pi$, have very similar periods. Thus, the asteroseismological models for both stars have similar characteristics, and the periods are associated with the same pulsational eigenmode, that is characterized\footnote{Each eigenmode is characterized by three numbers, the harmonic degree $\ell$, that represents the number of total nodal lines on the star surface, the azimuthal order $m$, which is the number of longitudinal nodal lines, and the radial order $k$, which represents the number of radial nodes \cite{unno:1989}. For non-rotating stellar models, as is the case in this work, $m=0$.} by $\ell = 1$ and $k = 2$. 
In this context, it is of great interest to obtain a bound to the axion mass from the measurement of the rate of change of period using a star of different characteristics of G117-B15A and R548. 
Moreover, analysing the actual hints of stellar cooling excesses, \cite{Giannotti:2015kwo} found that axion like particles, among other exotic particles, seem to be the most probable responsible for the stellar cooling excesses. The results of \cite{Giannotti:2015kwo}, in addition to the ubiquity of unknown systematic errors on the constraints determined by astrophysical considerations to $g_{ae}$, reinforces the great importance of having independent constraints.

DBV white dwarfs are hotter than the DAVs, as consequence, the impact of axions emission would be significantly higher.  
Furthermore, for the evolutionary stages of DBV stars, the cooling rate is 
larger than for the DAVs. As a result, DBV WDs have rates of change of period about two orders of magnitude greater than the DAVs \cite{2004A&A...428..159C}. 
However, due to the difficulty in finding stable periods in this type of stars, a measurement of the period drift for these stars was lacking. Fortunately, \cite{redaelli} have been able to 
estimate a measurement of 
$\dot \Pi$ of the DBV WD PG 1351+489 
whose value is $\dot \Pi = (2.0 \pm 0.9) \times 10^{-13}$ s/s and corresponds to the mode of higher observed amplitude and period $\sim$ 489 s.  
The aim of this paper is, first, to study the impact of axion emission on the evolutionary and pulsational properties DB WDs. In particular, we are interested in assessing the effect of axion emission on $\dot \Pi$. 
The second goal of this paper is to derive a new and independent bound on the axion-electron coupling constant, based on the recent estimations of $\dot \Pi$ in PG 1351+489. 

The paper is organized as follows. In section \ref{models} we describe briefly the input physics and numerical tools. In section \ref{effects} we discuss how axion emission impact on DB evolution and DBV pulsation properties. Section \ref{sec:constraints} is dedicated to present the bounds on $g_{ae}$ derived 
from the DBV star PG 1351+489. Finally, in section \ref{conclusions} we summarize our results and conclusions.

\section{Input physics and numerical tools}
\label{models}
The evolutionary models employed in this work were computed with the \texttt{LPCODE} stellar evolution code \cite{2003A&A...404..593A,althaus2005,2016A&A...588A..25M}. \texttt{LPCODE} has been used in a wide variety of studies of low-mass stars evolution, as the formation and evolution of different types of WDs \cite{2006A&A...454..845M,2009ApJ...704.1605A,2010ApJ...717..183R,2013A&A...557A..19A,2015A&A...576A...9A}. 
The pulsation calculations were computed with the adiabatic, stellar pulsation code \texttt{LP-PUL} \cite{2002Ap&SS.279..281C,pul}, which is coupled to the \texttt{LPCODE}. \texttt{LP-PUL} code has been used in many studies of non-radial pulsations of different types of stars, particularly, was extensively used to study pulsations 
on WDs and pre-WD stars \cite{2012MNRAS.420.1462R,2013ApJ...779...58R,2014JCAP...08..054C,2014A&A...569A.106C,2016arXiv160206355C}. 
The initial DB-WD models were obtained by \cite{2006A&A...454..845M}, as described in \cite{2009ApJ...704.1605A}, by calculating the complete stellar evolution from the Zero Age Main Sequence. 
The helium-rich envelope is the consequence of a born again episode, where almost all of the H content is violently burned (see e.g. \cite{2006A&A...454..845M}). After the born again episode, the H-deficient remnant evolves to the stage of PG 1159 stars, and finally to the WD domain. 
The WD cooling sequences computed in this work take into account thermal and chemical diffusion, and gravitational settling of chemical elements, as described in \cite{1969fecg.book.....B}. The diffusion coefficients of these processes are taken from \cite{paquette:1986}.  
The equation of state for the low-density regime is that of \cite{magni:1979}, while for the high-density regime we consider the treatment given in \cite{segretain-1994}. The conductive opacities are those of \cite{2007ApJ...661.1094C}.  
The neutrino emission processes taken into account are the photoproduction, pair-annihilation, Bremsstrahlung effect, and plasmon emission. This last one is the most relevant in the evolutionary stages of interest in this work (DBV instability strip), and the emission rate were taken form \cite{haft:1994}. The emission rates of the other neutrino processes were obtained from \cite{itoh:1996}. The DFSZ axion emission processes included are the Compton and Bremsstrahlung processes, being this last one the only relevant in the white dwarf stage. Axion emission is included as an additional energy sink term in the energy equation, following the prescriptions of \cite{nakagawa:1987} for Bremsstrahlung process, and \cite{raffelt:1995} for Compton emission.  
In particular, the axion Bremsstrahlung emission for multicomponent plasmas for the regime of strong ionic correlations is given by
\begin{equation}
	\epsilon_{a}=1.08\times 10^{23}\,\frac{g_{ae}^2}{4\pi}\,T_7^4\sum_j\frac{X_jZ_j^2}{A_j}\,F_j(T,\rho) \,\,[\mathrm{erg/g/s}]
\end{equation} 
where the sum is taken over all considered nuclear species, being $X_j$, $Z_j$ and $A_j$ their abundance per unit mass, atomic number and mass number respectively. $F_j$ are corrections for ionic correlation effects \cite{nakagawa:1987}, and are functions of the density ($\rho$) and temperature ($T$). $T_7$ is the temperature measured in units of $10^7$ K. 
For a more detailed description of how the emission of axion is included into the \texttt{LPCODE} see \cite{miller:2014}.

\section{Effects of axion emission on DB WDs}
\label{effects}
\subsection{Impact on the evolutionary properties}
\label{sec:impacto:evol}

During the hot stages of helium-rich WDs, cooling is dominated by neutrino and photon emission. Neutrino emission is dominant at the beginning of the WD-cooling track, but it becomes less relevant than photon emission by the time evolution has proceeded to the domain of the DBVs, at luminosities 
$\log L_{\gamma}/L_{\odot}\sim -1$, that correspond approximately to the luminosity characteristic of the instability strip of DBVs. With further evolution, neutrino emission ceases, and cooling is due to photon emission.  

\begin{figure}
\centering
\subfigure{%
  \label{fig:ax:nu:0.870}%
  \includegraphics[angle=0,width=.48\textwidth]{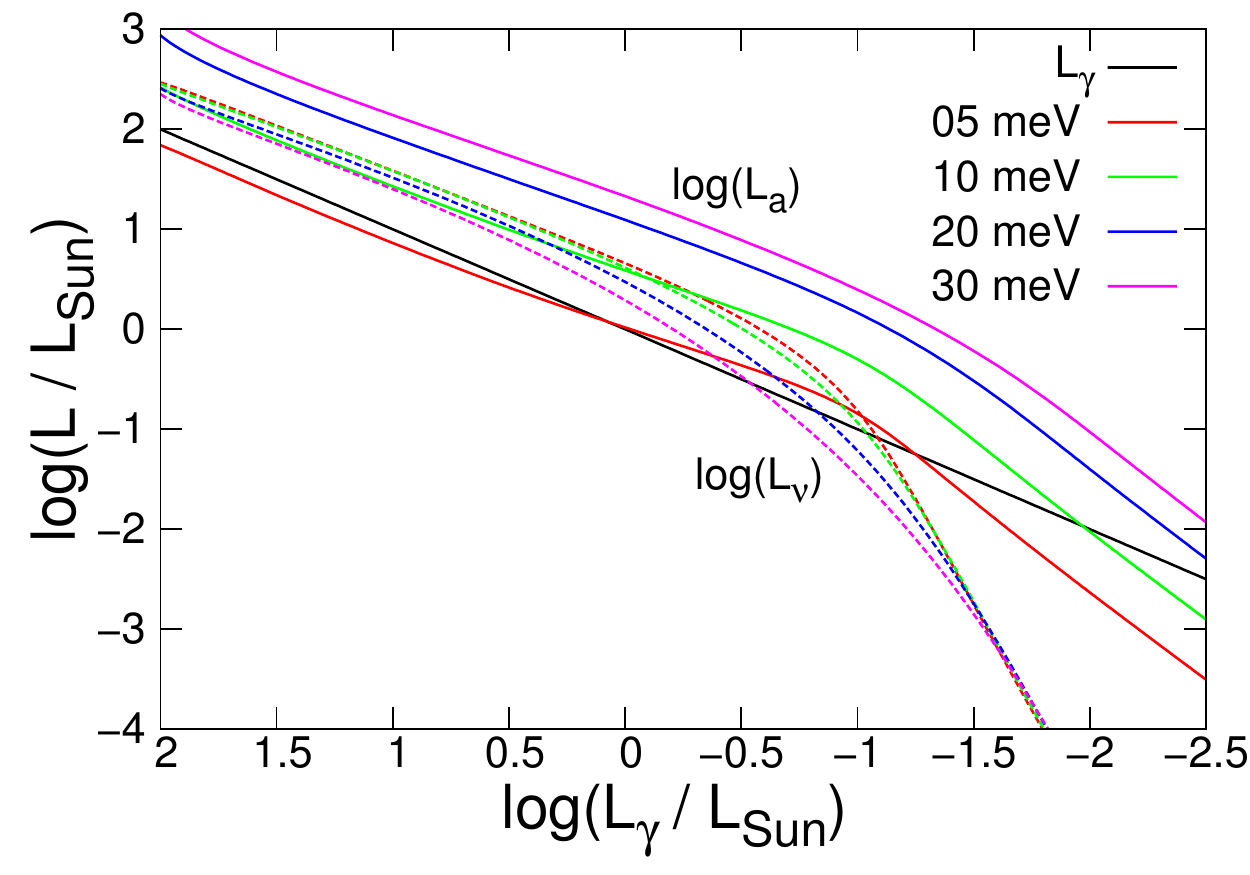}}
\subfigure{%
  \label{fig:lum:tpo:0.870}%
  \includegraphics[angle=0,width=.48\textwidth]{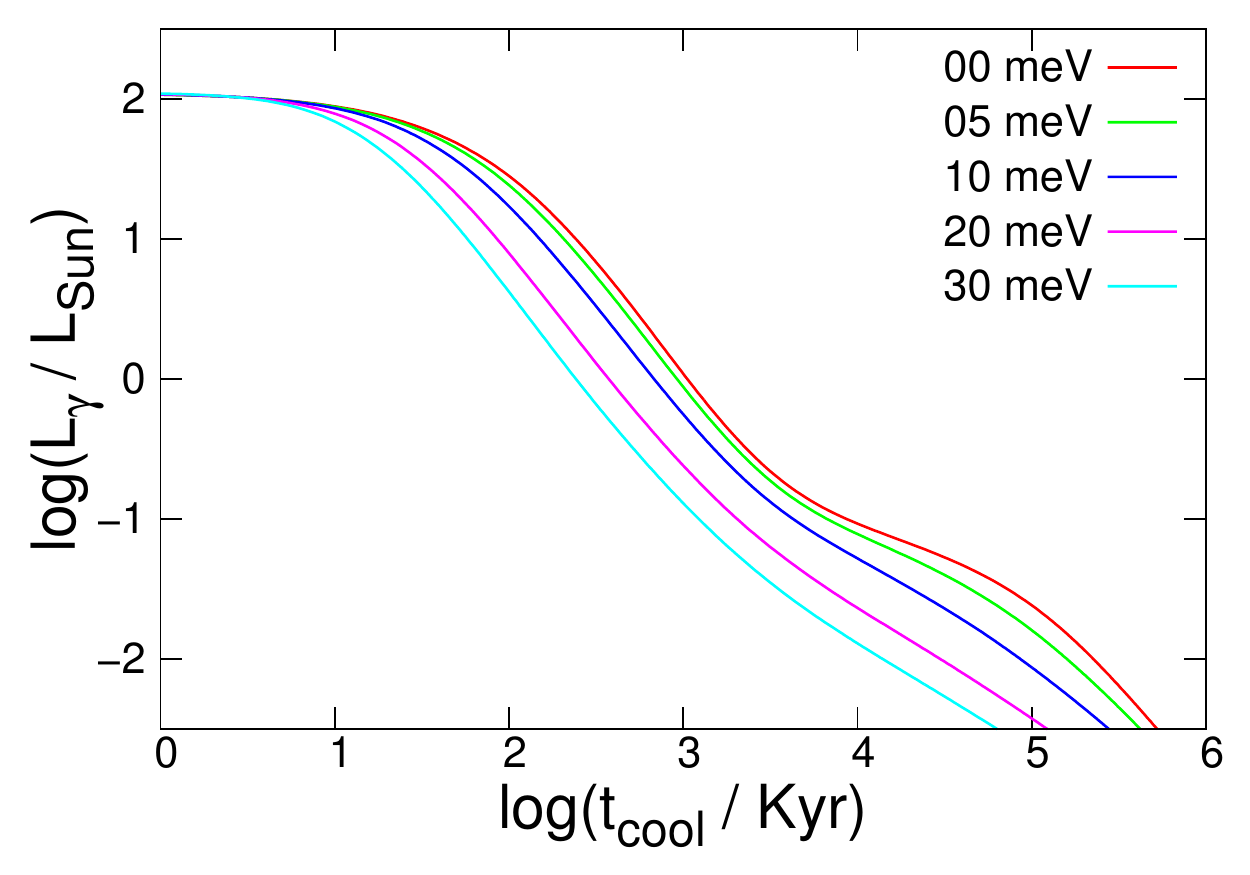}}    
\caption{Left panel: luminosity due to neutrino emission (dashed lines) and due to axion emission (solid lines) versus photon luminosity, for a WD model of 0.870 $M_{\odot}$ and for different axion masses. Right panel: luminosity due to photon emission versus the cooling time for the same WD model and different axion masses.}
  \label{fig:870}
\end{figure}

To assess the impact of axion emission on the evolutionary properties of DB WDs we computed evolutionary sequences for stellar masses of 0.515, 0.609 and 0.870$\,M_{\odot}$ and the following axion masses: 5, 10, 20 and 30 meV, thus covering the range of axion mass allowed by astrophysical considerations. We computed the WD evolutionary sequences in a self-consistent way, by taking into account the feedback of the axion emission into the thermal structure of the WD.  
As a result of this feedback, neutrino emission is altered, as found in \cite{miller:2014} for the case of DAVs. 
This is illustrated in the left panel of Figure \ref{fig:870} for our 0.870$\,M_{\odot}$ sequence, which shows that neutrino emission decreases with an increasing mass of the axion. 
This is because the production of neutrinos and axions takes place at approximately the same region in the WD interior.  
Axion emission deceases the inner temperatures of those regions in which they are produced, with the result that neutrino emission is markedly reduced, as compared with the situation in which axions are disregarded.  
However, the global effect is that, while neutrino emission decreases, the emission of axions still produces a net acceleration of the cooling. The resulting increase in the cooling rate becomes more important for larger axion masses (or coupling constant), as shown in the right panel of Figure \ref{fig:870}, where we depict the photon luminosity in terms of the cooling time for the sequences of $0.870\,M_{\odot}$ and different axion masses.  
   
\begin{figure}
\centering
  \includegraphics[angle=0,width=.99\textwidth]{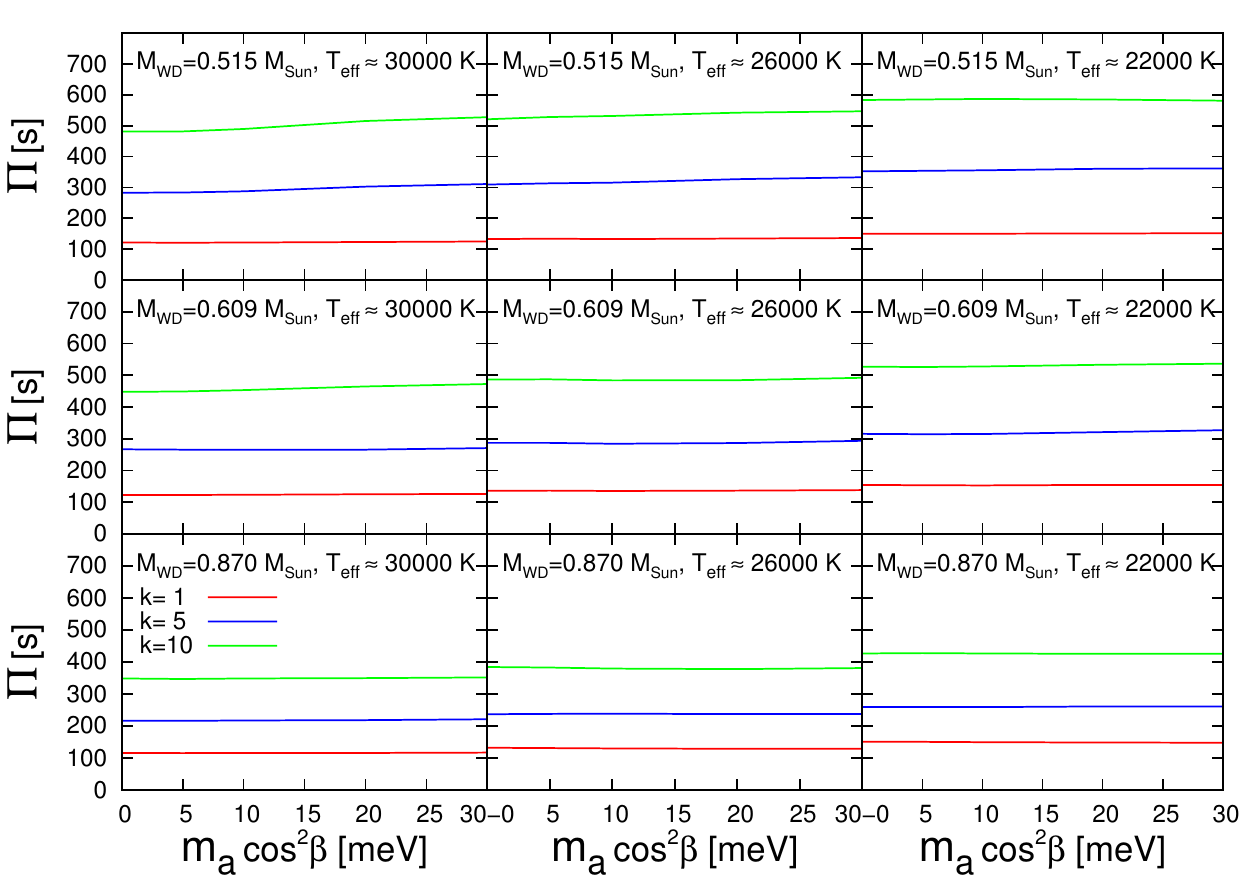}
  \caption{Pulsation periods versus axion mass corresponding to different WD masses, radial orders and effective temperatures ($T_{\mathrm{eff}}$) including the whole range of $T_{\mathrm{eff}}$ of the DBV instability strip.}
\label{fig:pe:gral}
\end{figure}

\begin{figure}
\centering
  \includegraphics[angle=0,width=.99\textwidth]{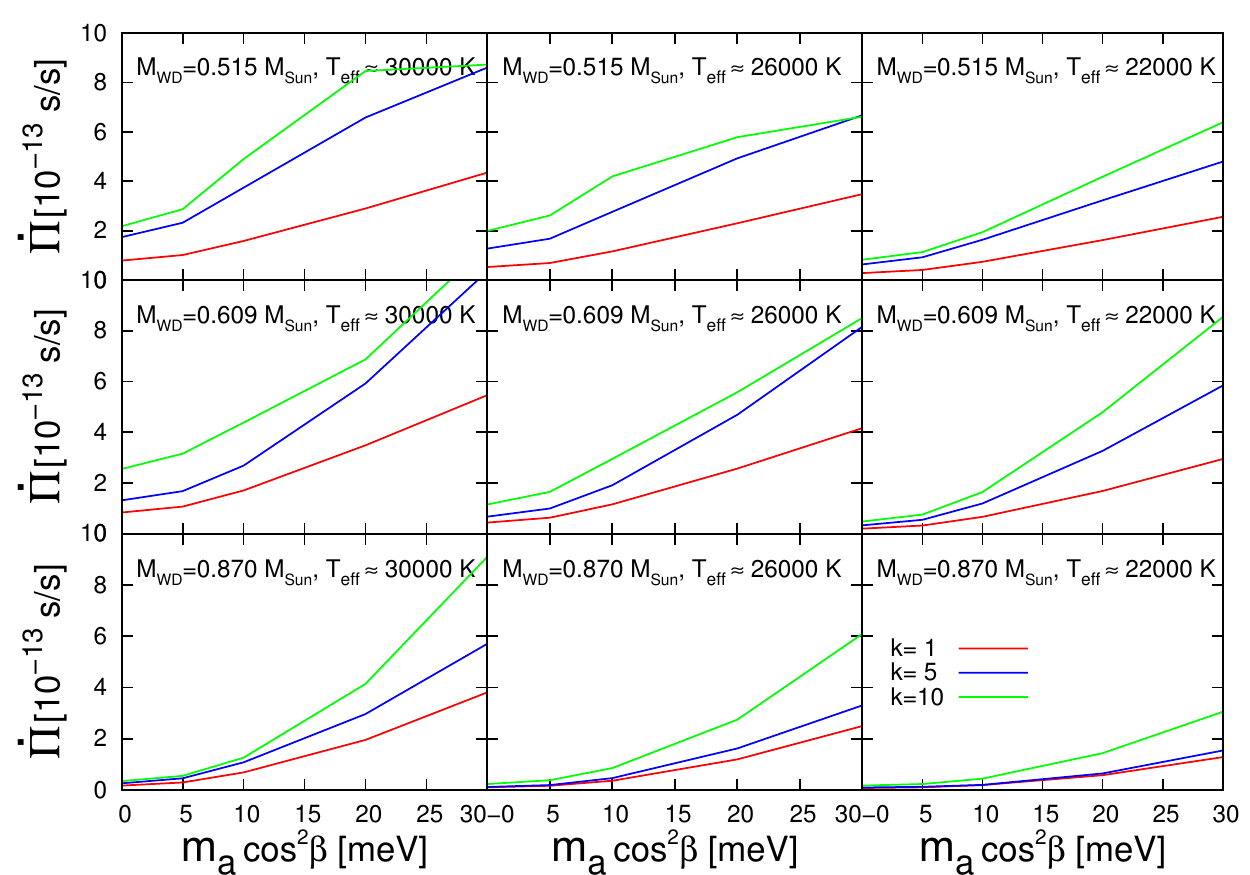}
  \caption{Same as in Figure \ref{fig:pe:gral}, but for the rates of period change.}
\label{fig:pp:gral}
\end{figure}

\begin{figure}
\centering
  \includegraphics[angle=0,width=.89\textwidth]{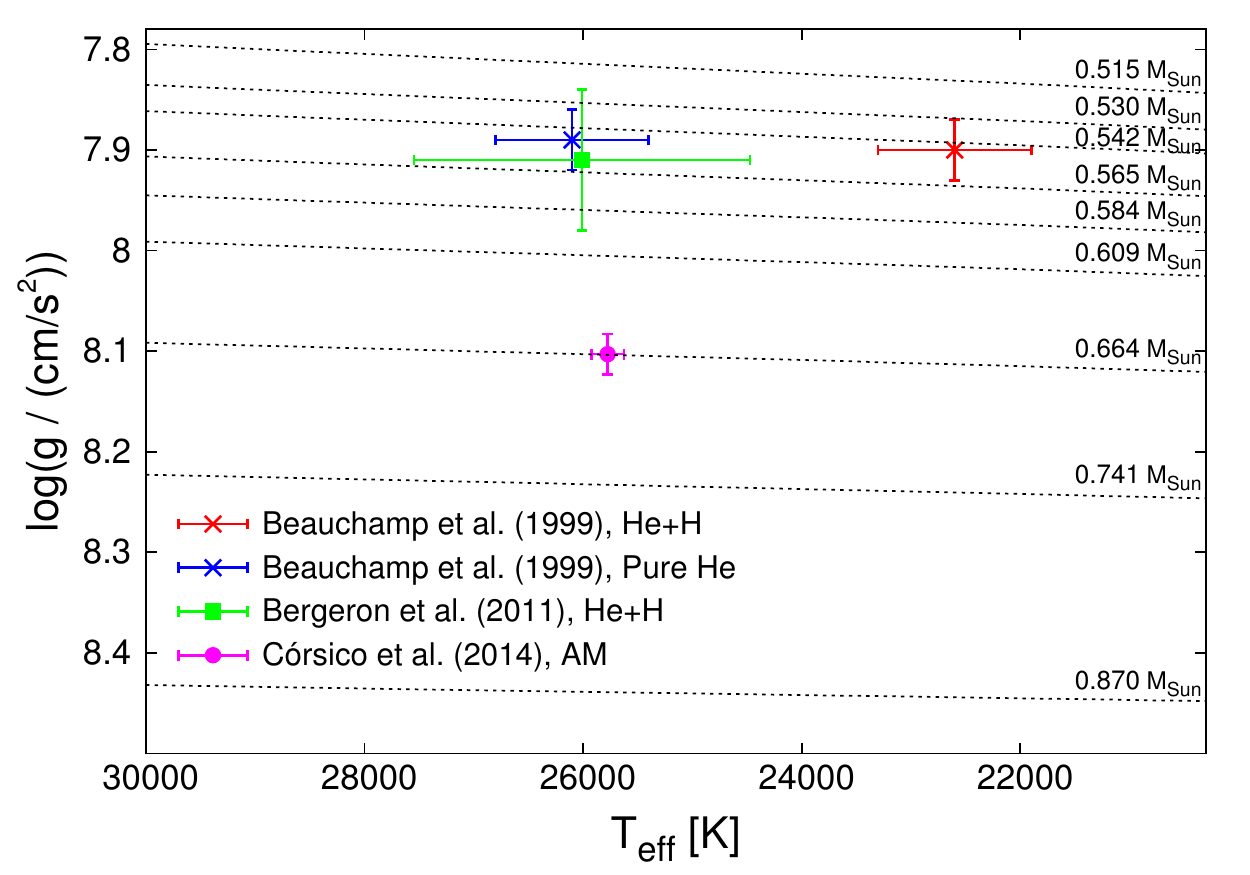}
  \caption{Effective temperature and surface gravity from the asteroseismological model of \cite{ale_estrellita} (pink circle), and the corresponding values derived spectroscopically for PG 1351+489 by different authors. The red (blue) cross correspond to an atmosphere of He with traces of H (pure He) and are from \cite{beauchamp:1999}. The green square correspond to an atmosphere with impurities of H and is from \cite{bergeron:2011}. The dashed black lines are the evolutionary DB-WD sequences for different stellar masses taken from \cite{2009ApJ...704.1605A}.}
\label{fig:logg-teff}
\end{figure}

\begin{figure}
\centering
  \includegraphics[angle=0,width=.79\textwidth]{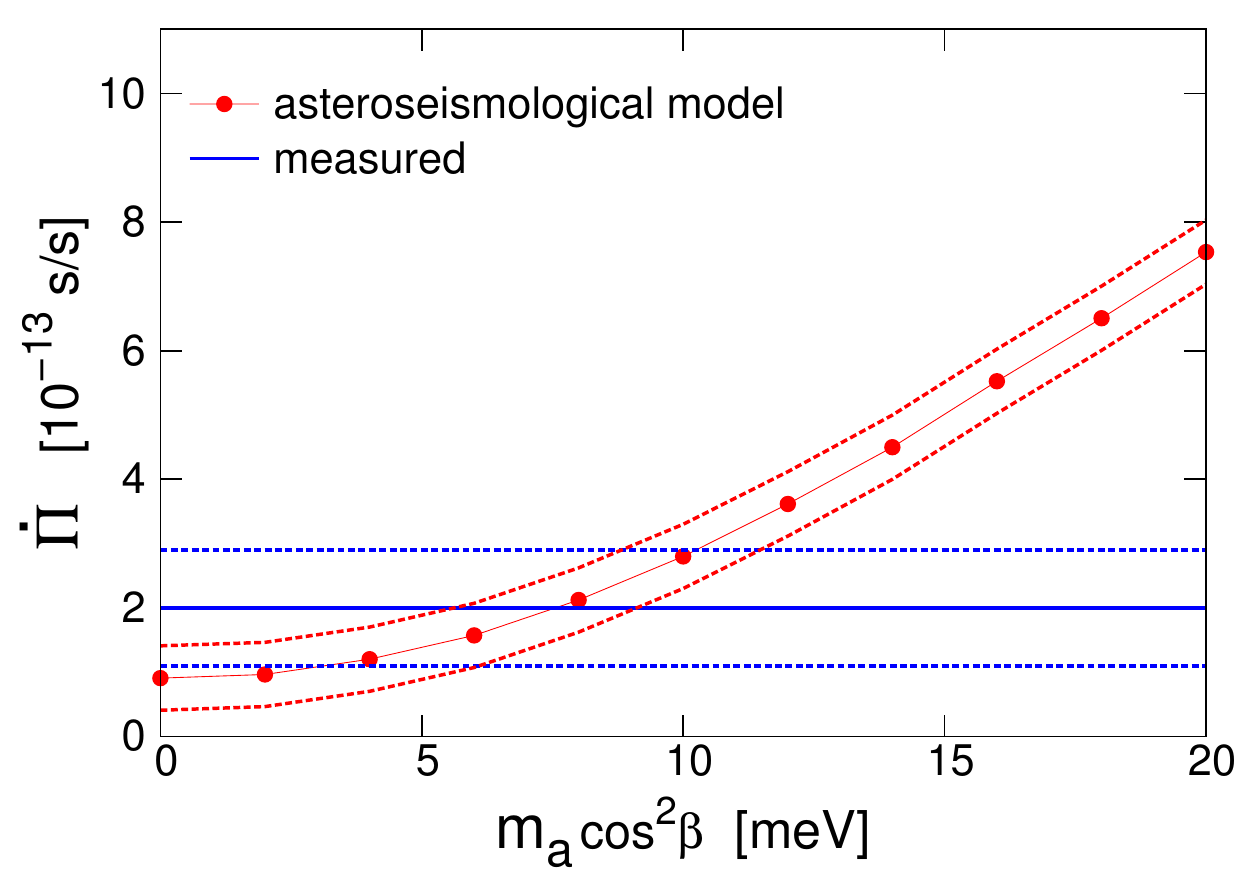}
  \caption{The rate of period change corresponding to the asteroseismological model for different values of $m_\mathrm{a}$ (red dots) with its error (dashed red lines). The value of $\dot \Pi$ estimated from the observations (filled blue line), and its error at 68\% CL (dashed blue lines).}
\label{fig:pp:cota}
\end{figure}

\begin{figure}
\centering
\subfigure{%
  \includegraphics[angle=0,width=.65\textwidth]{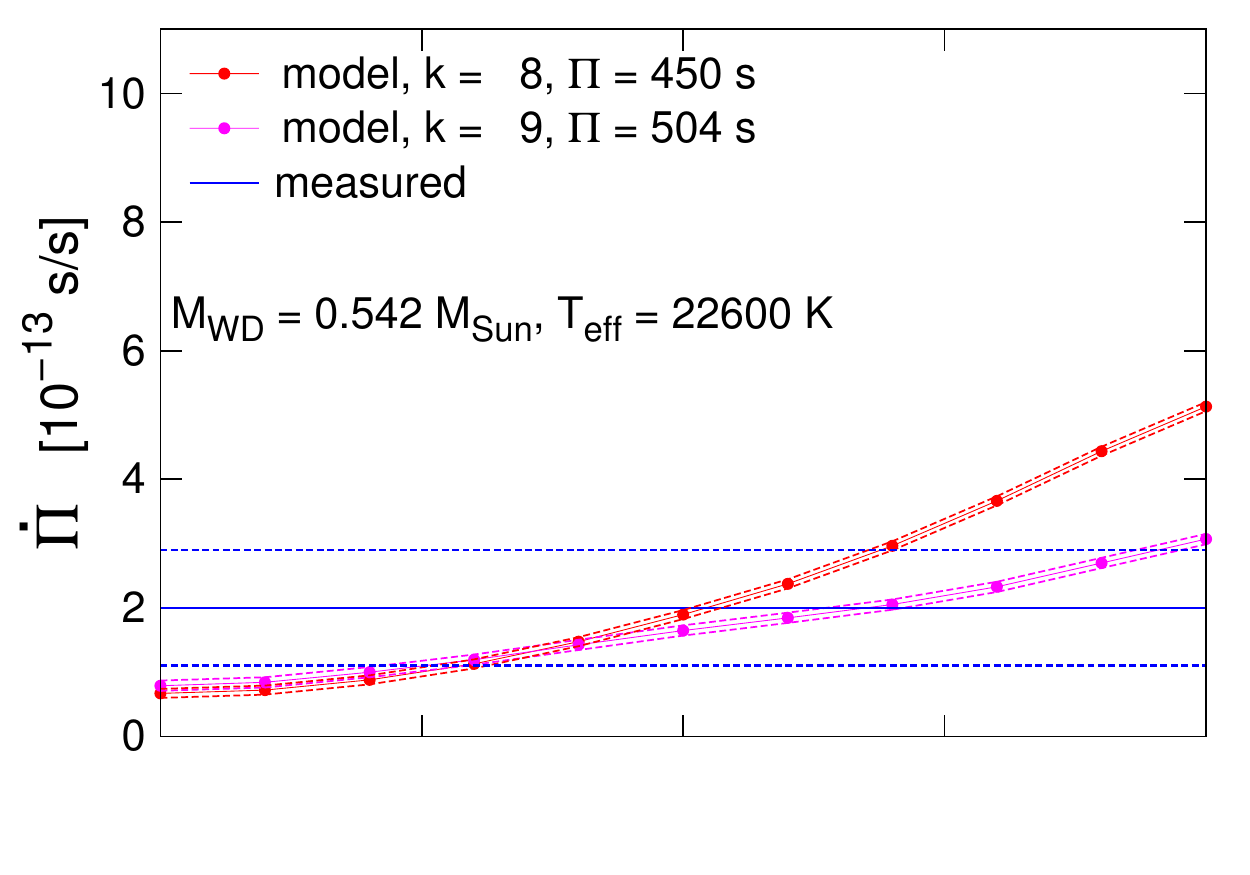}}
\label{fig:pp:542:22600}
\vspace{-1.6cm}
\subfigure{%
\centering
  \includegraphics[angle=0,width=.65\textwidth]{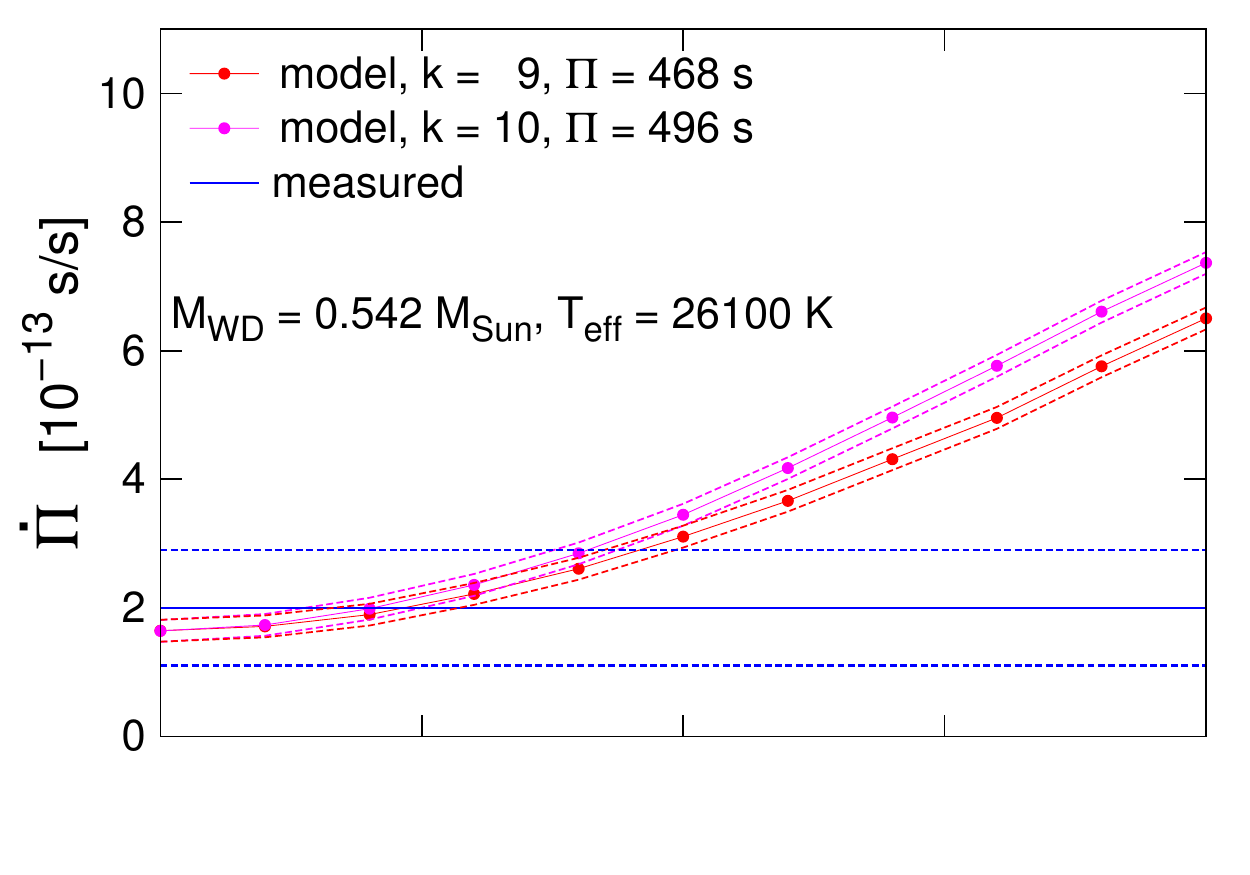}}
\label{fig:pp:542:26100}
\vspace{-1.6cm}
\subfigure{%
\centering
  \includegraphics[angle=0,width=.65\textwidth]{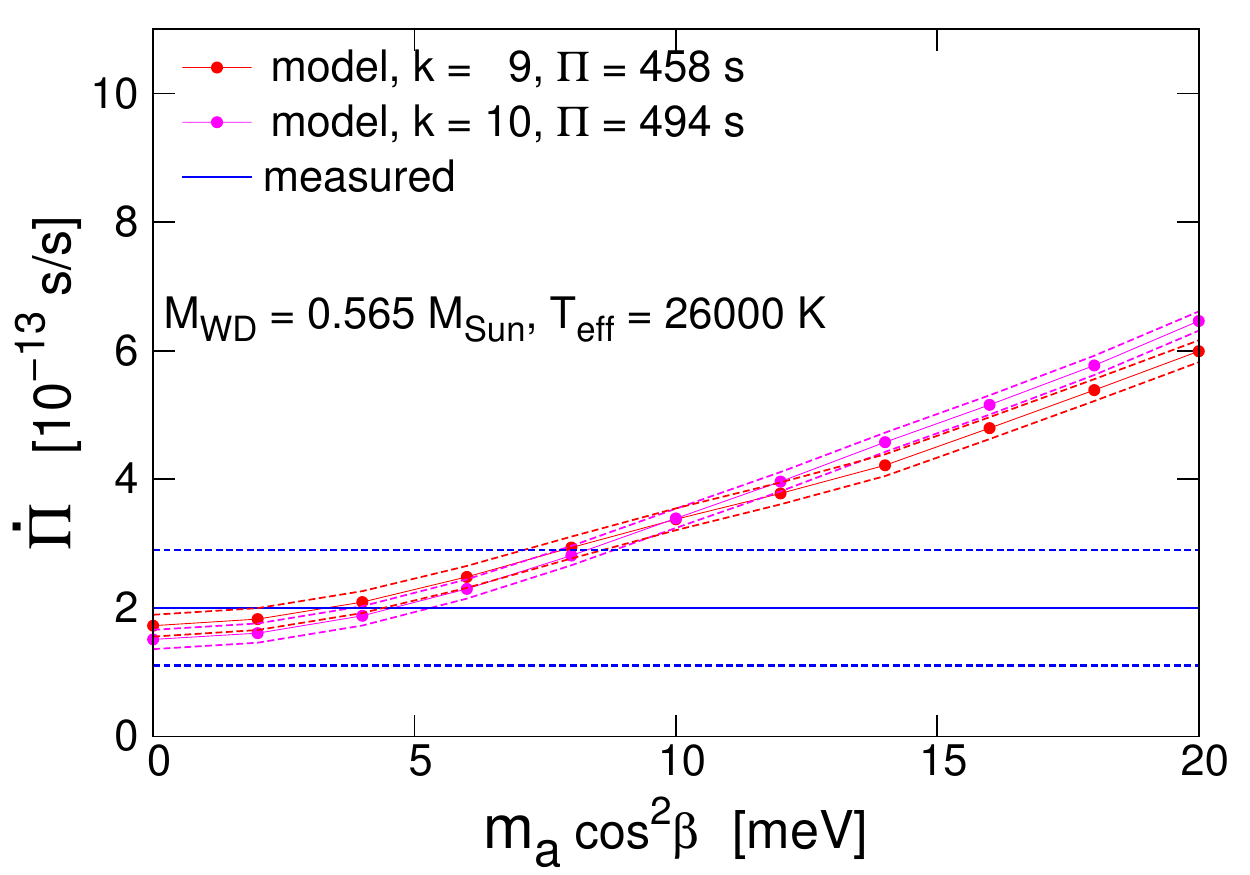}}
  \caption{The rate of period change for different values of the axion mass, for stellar models with $M_{\star}=0.542\,M_{\odot}$ and $T_{\rm eff}=22600\,$M (upper panel), and $T_{\rm eff}=26100\,$K (middle panel), and $M_{\star}=0.565\,M_{\odot}$ and $T_{\rm eff}=26000\,$K (lower panel). The red and pink dashed lines correspond to the estimated error taking into account the uncertainty in the $^{12}\mathrm{C}(\alpha,\gamma)^{16}\mathrm{O}$ reaction rate. The observed rate of change of period (and its error at 68\% CL) for PG1351+489 is depicted with filled (dashed) blue lines.}
\label{fig:pp:565:26000}
\end{figure}

\subsection{Impact on the pulsational properties}
\label{impact:pul}
The luminosity variations observed in WDs are due to non-radial $g$ modes with typical periods between 100 s and 1500 s. 
In order to study the effects of axion emission in the periods ($\Pi$) and the rate of change of periods ($\dot \Pi$) of the DBV stars, we computed the evolution of WD models for all the stellar masses and axions masses considered in section \ref{sec:impacto:evol} including the calculation of pulsations in the wide range of effective temperature characterizing the instability strip of the DBVs: $\sim 30000\,$K to $22000\,$K. 

Figure \ref{fig:pe:gral} shows the pulsation periods with harmonic degree $\ell=1$, radial order $k=1, 5$ and $10$, for effective temperatures of $22000\,$K, $26000\,$K and $30000\,$K, and WD masses of 
0.515, 0.609 and 0.870 $M_{\odot}$, in terms of $m_a \cos^2 \beta$. The variation in the periods due to axion emission, for a fixed effective temperature, is of 
 around $5 \%$. This is because the structure of the WD does not change significantly by considering axion emission. This result was first obtained by \cite{ale0} for the DAV WDs. 

For the case of the DBVs, since the effects of gravitational contraction in $\dot \Pi$ are negligible, the $\dot \Pi$ provides a measure of the cooling rate of the star. As cooling times are affected by axion emission, $\dot \Pi$ significantly increases when we consider axion emission, at variance with the negligible variations obtained for the periods. This result is consistent with the findings of \cite{ale0,ale,ale2} for DAV WDs.
In Figure \ref{fig:pp:gral} the effect of axion emission on $\dot \Pi$ is plotted for the same WD models and eigenmodes of Figure \ref{fig:pe:gral}.

\section{Constraints on axion mass from the DBV star PG 1351+489}
\label{sec:constraints}

As mentioned, in order to get a bound to the axion mass independently of that derived by asteroseismological studies of DAV stars, it would be of great interest to employ a measured value of $ \dot \Pi $ of a DBV WD. 
Fortunately, \cite{redaelli} obtained an estimate of the rate of change of a period  of a DBV WD, PG 1351+489, whose value is $ \dot \Pi = (2.0\pm 0.9) \times 10^{- 13} $ s/s and corresponds to the mode of larger observed amplitude with period $ \sim $ 489 s.  
PG 1351+489 is one of the 22 DBVs stars detected up to date \cite{2016MNRAS.457..575K,ale_estrellita}. 
Since the discovery of this star as a variable DB star \cite{winget:1987:PG}, it was considered as a good candidate to measure a value of a rate of period change. This is because of the presence of a single high-amplitude pulsation mode (that with a period of 489 s) in its power spectrum. In addition, there are three other normal pulsation modes of low-amplitude detected in this star \cite{alves:2003,redaelli}. 

The value obtained by \cite{redaelli} for $\dot \Pi$ was employed to place constraints to the neutrino magnetic dipole moment by \cite{ale_estrellita}, who found a detailed asteroseismological model for PG 1351+489 with a value of $\dot \Pi=0.81\times10^{-13}\,$s/s which is smaller respect to the observed one. 
In this work, we use the value of $\dot \Pi$ estimated by \cite{redaelli} to derive a limit for the axion-electron coupling constant, motivated by the results presented in section \ref{effects}, i.e., that the emission of axions is significant in the range of effective temperature characterizing the DBV instability strip, affecting in a considerable way the rates of change of period.
We follow the same procedure as in \cite{ale_estrellita}. First, we derive a bound fully relying on a detailed asteroseismological model for PG 1351+489. Second, we derive a less restrictive upper bound on the axion mass using models that take into account the wide range of effective temperatures derived spectroscopically for this star.

\subsection{Constraints from an asteroseismological model for PG 1351+489}

The asteroseismological model of \cite{ale_estrellita} for PG 1351+489 was obtained 
adopting the full evolutionary WD models computed by \cite{2009ApJ...704.1605A}, assuming $g_{\mathrm{ae}}=0$. The characteristics of the asteroseismological model are the following: stellar mass  $M_{\star}=0.664 \pm 0.013\,M_{\odot}$, effective temperature $T_{\rm eff}=25775 \pm 150\,$K, surface gravity $\log g = 8.103 \pm 0.020$ (in cgs) (see Fig. \ref{fig:logg-teff} ), and total helium (He) content $M_{\rm He}/M_{\star}\sim 5.5\times 10^{-3}$.  
In order to obtain models with different axion masses, we computed the WD evolution from the same model of $M_{\star}=0.664\,M_{\odot}$ for different axion masses. This was performed early in the WD evolution (at high luminosities) to allow the WD structure to relax to the new energy sink by the time 
it reaches the temperatures typical of DBV stars. The adopted values of axion mass were $m_a \cos^2 \beta$ from 0 meV to 20 meV in steps of 2 meV.  
At fixed effective temperature the periods do not vary significantly when axion emission is taken into account (sec. \ref{impact:pul}). Therefore, we constrain the axion mass by using the same asteroseismological model ($M_{\star}$, $T_{\rm eff}$, $M_{\rm He}$) that we considered for the case of $g_{\mathrm{ae}}=0$. 

The results are plotted in Figure \ref{fig:pp:cota}, where we show the values of $\dot \Pi$ for the different values of the axion mass. 
The uncertainty in the theoretical value ($\epsilon_{\dot\Pi}=0.5\times10^{-13}$ s/s) takes into account the coarseness of the mass grid adopted in the determination of the asteroseismological model. 
Below, we discuss briefly the theoretical uncertainties. We obtain a constraint for the axion mass of $m_a\cos^2 \beta \lesssim 11.5$ meV ($g_{ae} \lesssim 3.2\times 10^{-13}$) at the 68\% CL.

\subsubsection{Theoretical uncertainties}
The theoretical uncertainty of $\dot \Pi$ plotted in figure \ref{fig:pp:cota} 
was estimated in \cite{ale_estrellita} by comparing the value of $\dot \Pi$ of the asteroseismological model with those values corresponding to eigenmodes with similar periods belonging to models with $M_{\star}=0.741\,M_{\odot}$ and $M_{\star}=0.609\,M_{\odot}$ at the same effective temperatures than the asteroseismological model. These are the immediate lower an higher mass values of the grid (see Fig. \ref{fig:logg-teff}). 
Assuming that the uncertainties do not depend on $g_{ae}$, the error derived in \cite{ale_estrellita} was extrapolated to the values of $\dot \Pi$ with axion emission. 
Additional uncertainties in the theoretical quantities result from different uncertainties throughout the calculation of the previous stellar evolution. As the periods of a WD are sensitive to the adopted chemical profile, the uncertainties in the chemical profile translate into uncertainties in the structure of the asteroseismological model. 
The chemical profile of a WD depends on the $^{12}\mathrm{C}(\alpha,\gamma)^{16}\mathrm{O}$ reaction rate and extramixing adopted durign the core helium burning phase of the WD progenitor. 
Also, any process that affects the cooling rate of the WD could affects the value of $\dot \Pi$. For example, the exact values of the rate of neutrino emission and of the mass of the He envelope of the model influence the cooling times. 

A robust estimate of the theoretical errors due to the uncertainties in the previous
evolution of the WD is not a trivial issue, and is beyond the scope of
this work. However, in order to have a rough estimate of the impact of the uncertainties in the progenitor evolution, we estimated the changes in the value of $\dot \Pi$ for models with effective temperature, stellar mass and periods similar of the asteroseismological model, 
but with chemical profiles that have been artificially modified at high temperatures. 
The value of the $^{12}\mathrm{C}(\alpha,\gamma)^{16}\mathrm{O}$ reaction rate has an uncertainty of about $\pm 30$\% \cite{kunz:2002} and it affects directly the relative abundance of $^{12}$C and $^{16}$O of the resulting WD core. Since the specific heat of the carbon-oxygen (C-O) plasma is lower the higher the O content, 
a core made of pure O will cool faster than a mixture of C and O. 
Therefore, in the extreme case of a core made of pure O, the value of $\dot \Pi$ reached at the temperature of the asteroseismological model would be higher than in a mixture of C and O. 
Even in this case, we obtain a value of $\dot \Pi$ lower than the observed one. 
We obtain $\dot \Pi =1.19\times 10^{-13}$ s/s, which is $\Delta\dot\Pi_{\rm CO}=0.38\times 10^{-13}$ s/s higher than in the asteroseismological model. We conclude that, even in the extreme case in which the core is made of pure O, the variation $\Delta\dot\Pi_{\rm CO}$ is smaller than $\epsilon_{\dot \Pi}$. 
 
We also artificially modified the content of He of the model at high temperature. The total He content of the full-evolutionary model adopted in the asteroseismological model is  
$M_{\rm He}/M_{\star}\sim 5.5\times 10^{-3}$. At the luminosities corresponding of the DBV instability strip, WD models with a thinner He layer cool slightly faster than a model with a thicker He layer \cite{2004A&A...428..159C,1990ApJS...72..335T}. Therefore, we artificially reduced the He content by two order of magnitude. 
The value of $\dot \Pi$ reached at the temperature of the asteroseismological model is slightly higher than in the asteroseismological model. The difference is $\Delta\dot\Pi_{\rm He}=0.27\times 10^{-13}$ s/s, lower than $\epsilon_{\dot \Pi}$. In summary, the uncertainties in $\dot \Pi$ due to the uncertainties in the C-O chemical profile and in the value of $M_{\rm He}$ are included in the uncertainty $\epsilon_{\dot \Pi}$ adopted in the previous section to infer the upper limit to $g_{ae}$.

\subsection{Constraints taking into account spectroscopic data}

While the previous constraint relies on the accuracy of the asteroseismological model, it is also possible to obtain to the axion mass relying on the spectroscopically inferred properties and the theoretically predicted period changes of DBV stars.

The inferred effective temperature for PG 1351+489 depends on the adopted chemical composition in the model atmosphere calculations. The values of effective temperature and surface gravity derived by different authors are listed in Table \ref{tab:2}, and plotted in Figure \ref{fig:logg-teff} with the corresponding values from the asteroseismological model of \cite{ale_estrellita}. The values quoted in \cite{beauchamp:1999} correspond to an atmosphere of pure He, and an atmosphere with traces of H. The effective temperature derived by \cite{bergeron:2011} was obtained from a model atmosphere with impurities of H. 

\begin{table}[!ht]
\centering
\begin{tabular}{cccc}
\hline\hline\noalign{\smallskip}
Reference & Composition & $T_{\mathrm{eff}}$ [K] & $\log g$ [cgs]\\
\hline\noalign{\smallskip}
\cite{beauchamp:1999} & He + H &  22600 $\pm$ 700 &  7.90 $\pm$ 0.03\\
\cite{beauchamp:1999} & pure He  &  26100 $\pm$ 700 &  7.89 $\pm$ 0.03\\
\cite{bergeron:2011}  & He + H  &  26010 $\pm$ 1536 &  7.91 $\pm$ 0.07\\
\hline
\end{tabular}
\caption{Effective temperatures and surface gravities determined spectroscopically for PG 1351+489.}
\label{tab:2}
\end{table}

In order to obtain a bound to $g_{ae}$ almost independently of the asteroseismological model and its uncertainties, we compared the observed value of $\dot \Pi$ with those corresponding to different WD models, chosen in order to take into account the wide range of the spectroscopic effective temperatures derived for this star.  
The temperatures and masses of the models, and the characteristics of the modes employed in the analysis are listed in Table \ref{tab:3}. For each model we chosen two pulsation eigenmodes with periods that enclose the observed period of 489 s.

As the dominant uncertainty in the cooling rates of DBV stars is that introduced by the $^{12}\mathrm{C}(\alpha,\gamma)^{16}\mathrm{O}$ reaction rate \cite{2016arXiv160306666F}, we estimated an uncertainty in the theoretical values of $\dot \Pi$ due to the uncertainty in the value of the $^{12}\mathrm{C}(\alpha,\gamma)^{16}\mathrm{O}$ reaction rate. In the frame of the Mestel cooling law, \cite{1986ApJ...302..530K} obtained
\begin{equation}
	\dot \Pi=2\times10^{-30}A\left(\frac{\mu}{\mu_e^2}\right)^{0.286} \left(\frac{M_{\star}}{M_{\odot}}\right)^{-1.190}T_{\rm eff}^{2.857}\,\Pi
\end{equation}
where $A$ is the mass number and $\mu$ the mean molecular weight at the envelope. For a given period, the only quantity affected by the C/O ratio of the core is the mean mass number. 
We can write the change in $\dot \Pi$ due to the change in $A$ as
\begin{equation}
	\mathrm{d}\dot\Pi= \dot\Pi\,\frac{\mathrm{d}A}{A}.
\end{equation}
In the work of \cite{ale}, the authors changed the value of the reaction rate in a factor $f=0.5$ and $f=1.5$, and computed the previous stellar evolution for each value of $f$. The central abundances of C and O obtained by these authors for the resulting WD core are $X_{\rm C}=0.505$ and $X_{\rm O}=0.482$ respectively, for the case of $f=0.5$, and $X_{\rm C}=0.193$ and $X_{\rm O}=0.795$ for the case of $f=1.5$. Using these abundances we calculated a mean mass number for the C-O plasma for each value of $f$, $<A_{f=1.5}>$ and $<A_{f=0.5}>$, and estimated the relative change in $A$ as
\begin{equation}
	\frac{\mathrm{d}A}{A}=2\,\frac{<A_{f=1.5}>-<A_{f=0.5}>}{<A_{f=1.5}>+<A_{f=0.5}>}\sim0.087.
\end{equation} 
This implies an uncertainty in $\dot \Pi$ of $\epsilon_{\dot \Pi}\simeq 0.1\,\dot\Pi$. 
The resulting values of $\epsilon_{\dot \Pi}$
are listed in Table \ref{tab:3}, and are very small values.

The results of our analysis are shown in figure \ref{fig:pp:565:26000}. The more conservative constraint cames from the coolest model, and 
results $m_a\cos^2{\beta}\lesssim$ 19.5 meV, or for the axion-electron coupling constant $g_{ae}\lesssim 5.5 \times 10^{-13}$. 

\begin{table}[!ht]
\centering
\begin{tabular}{ccccccc}
\hline\hline\noalign{\smallskip}
Mass [$M_{\odot}$] & $g$ [cm/s$^2$] & $T_{\mathrm{eff}}$ [K] & $k$ & $\Pi$ [s] & $\dot \Pi$ [$10^{-13}$ s/s] & $\epsilon_{\dot \Pi}$ [$10^{-13}$ s/s]\\
\hline\noalign{\smallskip}
0.542 & 7.89 & 22600 & 8 & 450 & 0.69 & 0.07 \\
 & & & 9 & 504 & 0.79 & 0.08 \\
\hline\noalign{\smallskip}
0.542 & 7.88 & 26100 & 9 & 468 & 1.65 & 0.17 \\
 & & & 10 & 496 & 1.64 & 0.17 \\
\hline\noalign{\smallskip}
0.565 & 7.92 & 26000 & 9 & 458 & 1.17 & 0.17 \\
 & & & 10 & 494 & 1.51 & 0.15 \\
\hline
\end{tabular}
\caption{Temperatures, surface gravities, stellar masses and the characteristics of the modes employed to derive a limit on $g_{ae}$. $k$ is the radial order, $\Pi$ the pulsation period, $\dot \Pi$ its rate of change, and $\epsilon_{\dot \Pi}$ an estimated uncertainty that take into acoount the uncertainty in the $^{12}\mathrm{C}(\alpha,\gamma)^{16}\mathrm{O}$ reaction rate.}
\label{tab:3}
\end{table}

\section{Summary and conclusions}
\label{conclusions}

We studied the impact of axion emission on the evolutionary and pulsational properties of DB WDs, based on state-of-the-art evolutionary models of DB WDs consistent with the complete evolution of the progenitor stars. We covered the wide range of axion masses allowed by astrophysical considerations, and calculated non-radial, adiabatic g-mode pulsations in the range of effective temperatures characteristic of the inestability strip of DBVs. Our results indicate that at fixed temperatures the periods do not vary significantly with axion mass, while the $\dot \Pi$ are significantly affected. These results agree with the findings of \cite{ale0} for the case of DAV WDs. 

We also derived for a first time an upper bound to axion-electron coupling constant employing a DBV star, PG1351+489, using the estimation of $\dot \Pi$ for this star derived by \cite{redaelli}. 
Employing an asteroseismological model for this star obtained by \cite{ale_estrellita}, that reproduces the four periods of pulsation observed in PG1351+489, we obtained an upper bound of $g_{ae}<3.3\times10^{-13}$, or in terms of the axion mass $m_a\cos^2{\beta}\lesssim$ 11.5 meV. 
Based on DB WD models that reproduce the spectroscopic characteristics of PG1351+489, which are independent of a particular asteroseismological model, we obtained $g_{ae}<5.5\times10^{-13}$ or $m_a\cos^2{\beta}\lesssim$ 19.5 meV.
These results are less restrictive than the bounds derived by WDLFs, and can be improved if the uncertainties in the measured value of $\dot \Pi$ are improved. Based on the upper limit of the measured value of $\dot \Pi$ and taken into account the wide range of effective temperature derived for PG1351+489, the bound is consistent with that derived by \cite{ale} and \cite{ale2} from the asteroseismological study of DAV WDs.  It is worth noting that our results are independent of the limits derived from WDLFs, red giant stars on globular clusters and the asteroseismological study of DAV WDs, and therefore are affected by different unknown systematic errors.

\acknowledgments  
Part  of this work  was supported by AGENCIA through the Programa de Modernizaci\'on Tecnol\'ogica BID 1728/OC-AR, by the PIP 112-200801-00940 grant from CONICET, and by Proyecto de Incentivos (11/G110) grant by UNLP. M3B thanks the Alexander von Humbolt Foundation for a Return Fellowship. This research has made use of NASA Astrophysics Data System.
\bibliographystyle{JHEP}

\providecommand{\href}[2]{#2}\begingroup\raggedright\endgroup


\end{document}